\documentclass[conference]{IEEEtran}
\IEEEoverridecommandlockouts
\usepackage{cite}
\usepackage{amsmath,amssymb,amsfonts}
\usepackage{algorithmic}
\usepackage{graphicx}
\usepackage{textcomp}
\usepackage{xcolor}

\def\BibTeX{{\rm B\kern-.05em{\sc i\kern-.025em b}\kern-.08em
    T\kern-.1667em\lower.7ex\hbox{E}\kern-.125emX}}
\begin{document}

\title{Arbiter PUF: Uniqueness and Reliability Analysis Using Hybrid CMOS-Stanford Memristor Model\\}

\author{\IEEEauthorblockN{Tanvir Rahman}
\IEEEauthorblockA{\textit{Dept. of EEE} \\
\textit{BUET}\\
Dhaka, Bangladesh \\
1906176@eee.buet.ac.bd}
\and
\IEEEauthorblockN{A.B.M. Harun-ur Rashid}
\IEEEauthorblockA{\textit{Dept. of EEE} \\
\textit{BUET}\\
Dhaka, Bangladesh \\
abmhrashid@eee.buet.ac.bd}

}

\maketitle

\begin{abstract}
In an increasingly interconnected world, protecting electronic devices has grown more crucial because of the dangers of data extraction, reverse engineering, and hardware tampering. Producing chips in a third-party manufacturing company can let hackers change the design. As the Internet of Things (IoT) proliferates, physical attacks happen more, and conventional cryptography techniques do not function well. In this paper, we investigate the design and assessment of PUFs using the Stanford Memristor Model, utilizing its random filament evolution to improve security. The system was built using 45nm CMOS technology. A comparison is made between CMOS-based and memristor-based Arbiter PUFs, evaluating their performance under temperature, voltage, and process variations. Intra- and inter-hamming distances are employed by Monte Carlo simulations to estimate uniqueness and reliability. The results show that memristor-based PUFs offer better reliability than CMOS-based designs, though uniqueness needs further improvement. Furthermore, this study sheds light on the reasonableness of memristor-based PUFs for secure applications in hardware security.
\end{abstract}


\section{Introduction}
Electronic devices, driving rapid technological evolution, have reshaped how we learn, work, and communicate. At their core, integrated circuits (ICs) enable this progress, but their growing complexity makes ensuring security and proper functionality a challenging task. Although challenging, we must remember its importance, as these devices ultimately store not only our personal information but also proprietary data, making them an attractive target for hackers [1].
Hardware security is a broad term referring to the use of hardware components to ensure security throughout all stages of the integrated circuit (IC) supply chain. This includes processes such as silicon semiconductor collection, IC design, specification, outsourcing of IPs/ICs, and the after-life cycle of an IC [2]. In the context of the Internet of Things (IoT), hardware security is advantageous. The IoT is defined as the use of smart computing devices to connect physical items over the internet [3]. Hardware vulnerability in the IoT sector includes IP design modification, insertion of malicious contents in the form of hardware Trojan, Side channel leakage, insecure tamper mechanism, PCB design modification, etc.

\subsection{Physically Unclonable Function (PUF)}
A Physical Unclonable Function (PUF) is a hardware security module that leverages the inherent physical characteristics of integrated circuits (ICs) to generate unique and unpredictable signatures. It functions as a one-way mechanism, where a specific output is produced from a given input, but the input cannot be deduced from the output. The fundamental concept behind PUF technology is the variation that occurs in an IC characteristic due to process variations during manufacturing. Due to intrinsic process variations, the length, width, oxide thickness, and doping levels of transistor devices can vary. Even when the same mask and manufacturing process are used, each integrated circuit (IC) is unique, resulting in normal manufacturing variability. As a result, the electrical characteristics of a transistor may slightly differ when fabricated on different devices [4]. PUFs exploit inherent manufacturing process variations to generate unique hardware identifiers, making them reproducible, unique, unclonable, one-way, unpredictable, and tamper-evident [5]. PUFs are categorized in two major groups based on the fabrication: Silicon and Non-Silicon PUFs. Optical PUFs, Paper PUFs, Acoustic PUFs [22], Magnetic PUFs, etc. are Non-Silicon PUFs. In Silicon PUFs, there are two classes: Delay-based PUFs (RO PUF, Arbiter PUF [23], etc.) and Memory-based PUFs (SRAM PUF [24], Flip Flop PUF, Butterfly PUF [25], etc.).
The quality of a PUF is evaluated using specific metrics that determine its suitability for a given application. Two common metrics of the PUFs are the following [6]:

\begin{itemize}
\item Uniqueness: It represents the ability of a PUF to uniquely distinguish a particular chip among a group of chips of the same type. Hamming distance (HD) is used between a pair of PUF identifiers to evaluate uniqueness. If two chips, $i$ and $j$ ($i \ne j$), have $n$-bit responses, $R_i$ and $R_j$ respectively for the challenge $C$, the average inter chip HD among $k$ chips is defined as:
\begin{equation}
\textit{Uniqueness} = \frac{2}{k(k - 1)} \sum_{i=1}^{k-1} \sum_{j=i+1}^{k} \frac{HD(R_i, R_j)}{n} \times 100\%
\end{equation}

\item Reliability: 
The reliability of the PUF captures how efficient a PUF is in reproducing the response bits. 
Intra-chip Hamming Distance (HD) is determined among several samples of PUF response bits to evaluate it. 
To estimate the HD in the chip, an \( n \)-bit reference response \( R_i \) is extracted from chip \( i \) in normal operating conditions (at room temperature using normal supply voltage). 
The same \( n \)-bit response is extracted at different operating conditions (such as different ambient temperature or supply voltage), denoted as \( R_i' \). 
A total of \( m \) samples of \( R_i' \) are collected. 
For chip \( i \), the average intra-chip HD is estimated as follows:
\begin{equation}
HD_{\text{INTRA}} = \frac{1}{m} \sum_{t=1}^{m} \frac{HD(R_i, R_{i,t}')} {n} \times 100\%
\end{equation}
Where \( R_{i,t}' \) is the \( t \)-th sample of \( R_i' \). 
\( HD_{\text{INTRA}} \) indicates the average number of unreliable PUF response bits. 
So, the reliability of a PUF can be defined as:
\begin{equation}
\textit{Reliability} = 100\% - HD_{\text{INTRA}}
\end{equation}
\end{itemize}

\subsection{Memristor}
Leon Chua, a circuit theorist, introduced the term “Memristor” in 1971 to describe the fourth fundamental circuit element. A memristor (short for memory resistor) is a nonlinear resistor with memory. Unlike resistors, capacitors, and inductors that define relationships between pairs of the four fundamental circuit variables (electric current ($i$), voltage ($v$), charge ($q$), and magnetic flux ($\phi$)), memristors correlate charge and flux. Chua proposed this fourth circuit element, defined by a ($\phi$-$q$) curve, as a two-terminal element providing a functional relationship between charge and flux, $d\phi = M\,dq$. When $M$ is constant, memristance behaves like resistance in linear elements. However, when $M$ varies with $q$, it results in a nonlinear element. A memristor, unlike a resistor, remembers its states or history. 
If memristor circuit is analogized to a hydraulic system, wire is like a pipe, current is water flowing in the pipe, voltage is the pressure controlling the water and memristor is like sand filter that catches sand as it flows through the water. Memristor can be viewed as two variable resistors in series, with resistances constant over the device’s length. If $D$ is the total width of the device, the equivalent resistance of the device is given by:  
\begin{equation}
R_{eq}(t) = \frac{w(t)}{D} \cdot R_{on} + \left(1 - \frac{w(t)}{D} \right) \cdot R_{off}
\end{equation}
where $R_{on}$ is the low resistance of the device if the entire device is doped, and $R_{off}$ is the high resistance of the device if the entire device is undoped. The instantaneous value of $w(t)$ depends on the history of the applied voltage. High-resistance state (HRS) and low resistance state (LRS) are defined by undoped and doped part of the film inside the memristor. And these doping and undoping processes caused by wire formation with vacancy migration, are unpredictable which causes randomness [7]. 
\subsection{Memristive Arbiter PUF}
Memristors exhibit process variations, stochastic switching, and non-volatility. By using these properties, memristive PUFs offer higher entropy, lower power consumption, smaller footprint, better resilience. Main criteria is geometry of the conductive filament within a device. It can switch from device-to-device to cycle-to-cycle due to the unpredictable nature of defect formation and elimination in the oxide material [21]. 

Several configurations of memristor-based implementations in Arbiter PUFs have been proposed by researchers to enhance performance. The authors of [8] introduced a challenge-dependent stage delay PUF consisting of two parallel memristor-based delay lines. Each stage in these delay lines is grounded through MOSFETs, which are controlled by two distinct challenge bits. At the end of the delay lines, a D flip-flop arbiter, initially set to logic 1, determines which of the two signals propagates faster through the circuit. In [9], the authors designed a PUF using a parallel string of memristors, modifying the previous architecture from [8]. They repositioned the CMOS challenge programming transistor from its original connection to the ground, placing it parallel to the memristors to enhance resistance against cryptanalysis attacks. The circuit functions in two phases: challenge application and response generation. During the challenge application phase, the control signal (Ctrl) remains high, preventing voltage buildup at the arbiter’s input terminals. In the response generation phase, Ctrl is set low, allowing voltage pulses to propagate through both delay paths. In [10], the number of response bits was increased by adding RS NAND latches. In [11], Teo et al. proposed an improved APUF similar to the one shown in [10] (except for the number of memristors per transistor and the number of challenge-response bits, e.g., 1-5 memristors per transistor and 8, 16, or 32-bit challenge to 4 or 8-bit response). They calculated the uniqueness, uniformity, and bit-aliasing of the improved APUF using SilTerra’s 180nm at 1.8V and 130nm at 1.2V, and the Biolek memristor model.

In this paper, we first presented the Stanford memristor model as the foundation for our design approach. We implemented a CMOS–memristor hybrid Arbiter PUF circuit, inspired by the architecture proposed in [9]. However, unlike [9], which utilizes the Biolek memristor model, our implementation adopts the Stanford model to capture more realistic device-level stochasticity. The Stanford model accounts for variations in both the tunneling gap and the conductive filament radius, which contribute to the intrinsic randomness essential for PUF functionality. These physical variations are harnessed in our proposed design, implemented using the GPDK 45 nm technology node within the Cadence Virtuoso environment. 

The remainder of the paper is organized as follows: Section II describes the mathematical representation of the Stanford memristor model, design and implementation of the arbiter PUF, detailing the circuit schematic, specifications, and the design parameters used for variation estimation, considering both CMOS and memristor technologies. Section III provides a series of figures that illustrate the effects of variations on the response. Additionally, Monte Carlo simulations are presented, showing the distribution of the average response value under process variation and mismatch. Section IV analyzes the results obtained from the study, focusing primarily on the key PUF metrics of reliability and uniqueness. The findings are analyzed in detail to assess the performance of the designed PUF. Additionally, a comparative analysis is conducted, where the obtained results are evaluated against those reported in existing research papers. Section V wraps up the study by summarizing the key findings and their importance. It also discusses the limitations of the work and areas that could be improved. 

\section{Design}

\subsection{The Mathematical Representation of Stanford Model}

The Stanford memristor model [12], developed by Li et al., captures key characteristics such as stochastic switching behavior, multi-level cell capability, switching voltage variation, and resistance distribution. $TiN/HfO_x/TiO_x/Pt$ bi-layer RRAM devices of 10 nm feature sizes were fabricated. Generation and recombination of oxygen vacancies in the oxide layer mainly maintain the conductive filament (CF) geometry. Tunneling gap distance ($g$) and CF radius ($r$) are the key control variables.

During the switching operation, the resistance of the conductive filament (CF) and the hopping current density within the gap both affect the voltage across the gap region. In the SET process, the CF undergoes growth in both length and radius. Conversely, during the RESET process, the release of $O_{2}^-$ ions from the electrode and their subsequent recombination control the evolution of the CF. The conduction behavior of the RRAM cell is modeled based on two primary mechanisms: hopping conduction paths and metallic conduction paths. A key feature leveraged in the proposed PUF design is the random variation effect. Specifically, the variations in low resistance state ($R_{LRS}$) and high resistance state ($R_{HRS}$) arise from stochastic fluctuations in the CF radius ($r$) and the tunneling gap ($g$), respectively: 
\begin{equation}
g = \int \left( \frac{dg}{dt} + \delta g \times \chi(t) \right) dt
\end{equation}
\begin{equation}
r = \int \left( \frac{dr}{dt} + \delta r \times \chi(t) \right) dt
\end{equation}
Here, \( \chi(t) \) represents a zero-mean Gaussian sequence with a root mean square (RMS) value of unity. The parameters \( \delta g \) and \( \delta r \) denote the variation amplitudes, which are determined based on empirical device measurement data.
\subsection{Design of Arbiter PUF}
Mathew et al. [8] proposed a memristor-based delay PUF with two parallel delay lines, where each stage is grounded via MOSFETs controlled by challenge bits (Figure 1). Chatterjee et al. [9] enhanced this by repositioning the CMOS challenge transistor parallel to the memristors, improving resistance to cryptanalysis.
We have adapted the circuit proposed in [9] to improve the security and achieve more stable PUF performance. The Biolek model is replaced by the Stanford model. Figure 2 shows the circuit. 
\begin{figure}[htbp]
\centering
\includegraphics[width=0.48\textwidth]{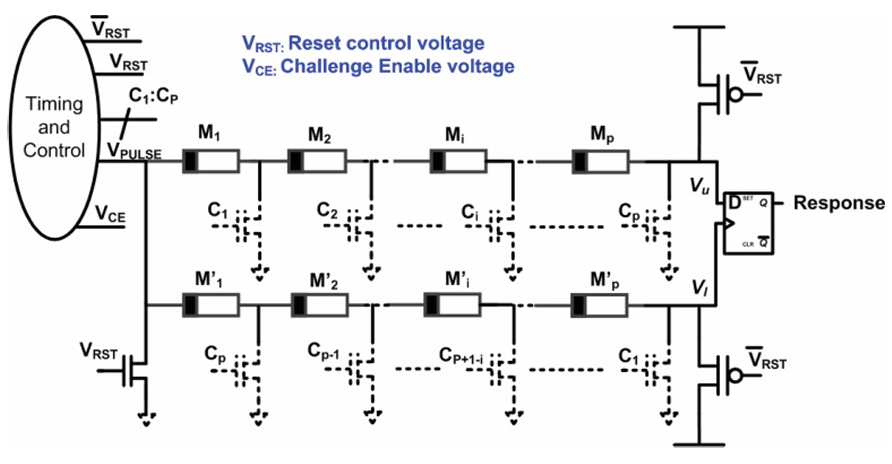}
\caption{Memristor-based Arbiter PUF circuit proposed by Mathew et al. [8]}
\label{fig1}
\end{figure}
\begin{figure}[htbp]
\centering
\includegraphics[width=0.48\textwidth]{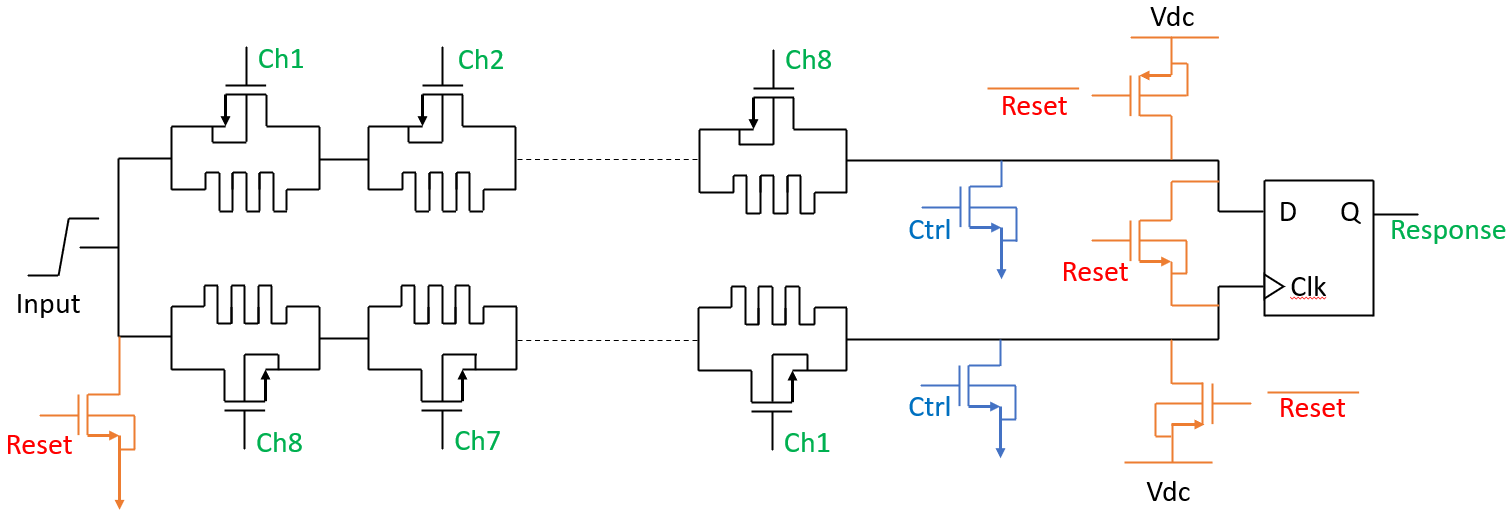}
\caption{Memristor-based Arbiter PUF for single response}
\label{fig2}
\end{figure}
The circuit operation consists of two main phases: reset and challenge-response. When $Reset=1$, a reset is performed. Due to varying effective potential differences across each memristor, they are influenced by their individual device-level properties and settle into random resistance states. In the challenge-response phase, when $Ctrl=1$, it prevents voltage accumulation at the arbiter’s input. Current flows across all memristor-NMOS parallel combinations and each challenge affects memristor resistance effectively. When $Ctrl=0$, voltage pulses travel along the delay paths, and response is produced according to the conventional PUF mechanism.
\begin{table}[htbp]
\renewcommand{\arraystretch}{1.05}
\setlength{\tabcolsep}{16pt}
\centering
\caption{Design Parameters and Nominal Values Used for Reliability Evaluation}
\label{table1:design_parameters}
\begin{tabular}{|c|c|}
\hline
\textbf{Design Parameter} & \textbf{Nominal Value} \\
\hline
Temperature (CMOS) & $27\,^\circ\mathrm{C}$ \\
\hline
Temperature (Memristor) & $27\,^\circ\mathrm{C}$ \\
\hline
Supply Voltage & 5~V \\
\hline
\end{tabular}
\end{table}
Intra-chip HD (Hamming Distance) is required for reliability estimation. First, a reference response was obtained from the device under nominal operating conditions. We then regenerated the response under different operating conditions and computed the intra-HD between the two. Table I shows the design parameters and their nominal values used in reliability test.
We used the Stanford memristor model in another circuit proposed in [10], where response bits are extracted from various stages in the delay paths. In this design, from 8-bit challenge, a 4-bit response is generated. The circuit is illustrated in Figure 3. 
\begin{figure}[htbp]
\centering
\includegraphics[width=0.48\textwidth]{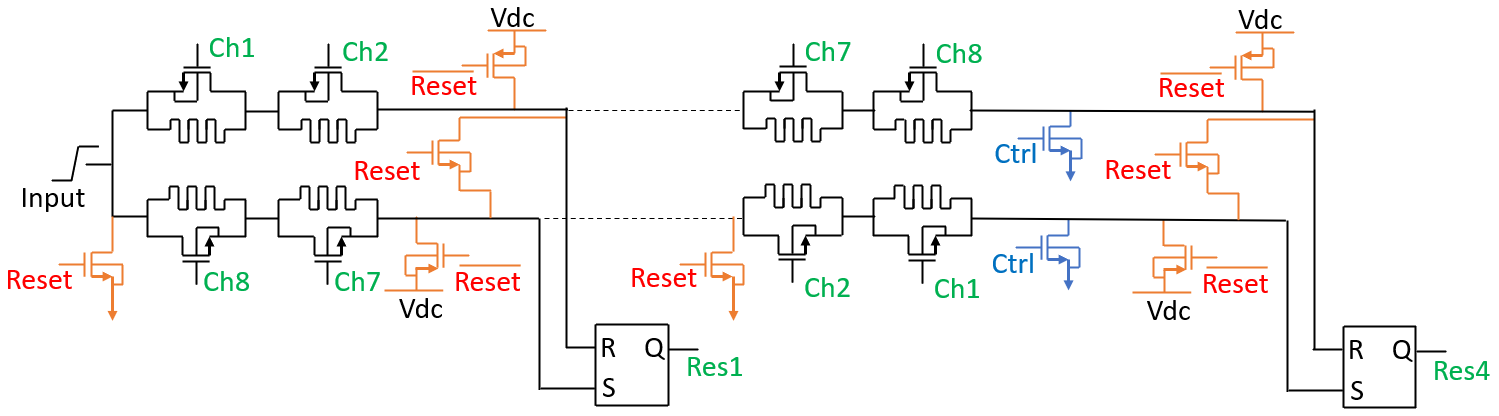}
\caption{Memristor-based Arbiter PUF for multiple response}
\label{fig3}
\end{figure}

\begin{table}[htbp]
\renewcommand{\arraystretch}{1.3}
\setlength{\tabcolsep}{6pt}
\centering
\caption{Design Parameters and Nominal Values for Single-Response Uniqueness Evaluation}
\label{table2:single_response_uniqueness}
\begin{tabular}{|c|c|c|}
\hline
\textbf{} & \textbf{Design Parameter} & \textbf{Nominal Value} \\
\hline 
& Doped Length & 3~nm \\
\cline{2-3}
& Filament Length & 3~nm \\
\cline{2-3}
{\centering \textbf{Memristor}} & Filament Width & 0.5~nm \\
\cline{2-3}
& Characteristics Tunneling Length & 0.4~nm \\
\cline{2-3}
& Gap Region Hopping Current Density & $1 \times 10^{13}$~A/m$^2$ \\
\hline
{\centering \textbf{CMOS}} 
& Process Corners (ff, fs, sf, ss) & tt \\
\hline
\end{tabular}
\end{table}
Uniqueness is estimated from nanoscale manufacturing process variations. From a circuit design perspective, these process variations are typically categorized into two main types: die-to-die variations and within-die variations [13]. Within-die variations occur among different circuit elements on the same die and can be categorized as either systematic or random. Systematic variations, with a radius of a few millimeters, arise from imperfections in the fabrication process and exhibit spatial correlation. Random variations, however, are non-systematic and result from factors such as random dopant concentration in transistors and gate oxide thickness variations. These are intrinsic to the silicon material and cannot be controlled. The radius of random variations is comparable to the size of individual devices, allowing each device to vary independently. These random variations are the primary source of unpredictable and unclonable responses in a PUF [14]. We utilized both memristor and CMOS device parameters to evaluate uniqueness. Table II shows the memristor and CMOS device parameters along with their nominal values used for a single response, while Table III summarizes the parameters for multiple responses.

\begin{table}[htbp]
\renewcommand{\arraystretch}{1.4}
\setlength{\tabcolsep}{10pt}
\centering
\caption{Design Parameters and Nominal Values for Multiple-Response Uniqueness Evaluation}
\label{table3:multi_response}
\scriptsize
\begin{tabular}{|c|c|c|}
\hline
\textbf{} & \textbf{Design Parameter} & \textbf{Nominal Value} \\
\hline
& Adjacent Oxygen Vacancy Distance & 0.25~nm \\
\cline{2-3}
& Oxygen Atom Vibration Frequency & $1 \times 10^{13}$~Hz \\
\cline{2-3}
& Average Active Energy for O Vacancy & 0.7~eV \\
\cline{2-3}
& Enhancement Factor in $E_a$, $E_h$ & 0.75~nm \\
\cline{2-3}
\textbf{Memristor} & Characteristics Tunneling Length & 0.4~nm \\
\cline{2-3}
& Hopping Current Density & $1 \times 10^{13}$~A/m$^2$ \\
\cline{2-3}
& Initial Gap Region Length & 3~nm \\
\cline{2-3}
& Filament Length & 3~nm \\
\cline{2-3}
& Filament Width & 0.5~nm \\
\cline{2-3}
& Gap Distance Amplitude & $4 \times 10^{-5}$ \\
\hline
\textbf{CMOS}
& Process Corners (ff, fs, sf, ss) & tt \\
\hline
\end{tabular}
\end{table}

\section{Results}
It is mathematically derived through circuit analysis in [8] that the response for a given challenge \( \vec{C} \) is determined by the sign of a scalar product between two vectors, and is given by:
\begin{equation}
\text{Response} = -\text{sgn}\left( \vec{w}^{T} \left( \vec{C} \right) \cdot \vec{\Phi} \right)
\end{equation}
where, we assume bipolar encoding for the response (i.e., logic-0 is represented by \(-1\)), and the vectors \( \vec{w}(\vec{C}) \) and \( \vec{\phi} \) are given by:
\begin{equation}
\vec{w}(\vec{C}) = 
\begin{pmatrix}
\mathrm{CMOS} \cdot \Delta M_1(\vec{C}) \\
\mathrm{CMOS} \cdot \Delta M_2(\vec{C}) \\
\vdots \\
\mathrm{CMOS} \cdot \Delta M_p(\vec{C})
\end{pmatrix}
\end{equation}
\begin{equation}
\vec{\Phi} = 
\begin{pmatrix}
n \\
n-1 \\
\vdots \\
1
\end{pmatrix}
\end{equation}
\noindent
where, $n$ is the number of challenge bits of the APUF; $\mathrm{CMOS}$ is the drain capacitance of each NMOS switch; and $\Delta M_i$ is the difference of resistances of the $i$-th memristors in the two (upper and lower) branches.
\begin{figure}[htbp]
\centering
\includegraphics[width=0.48\textwidth]{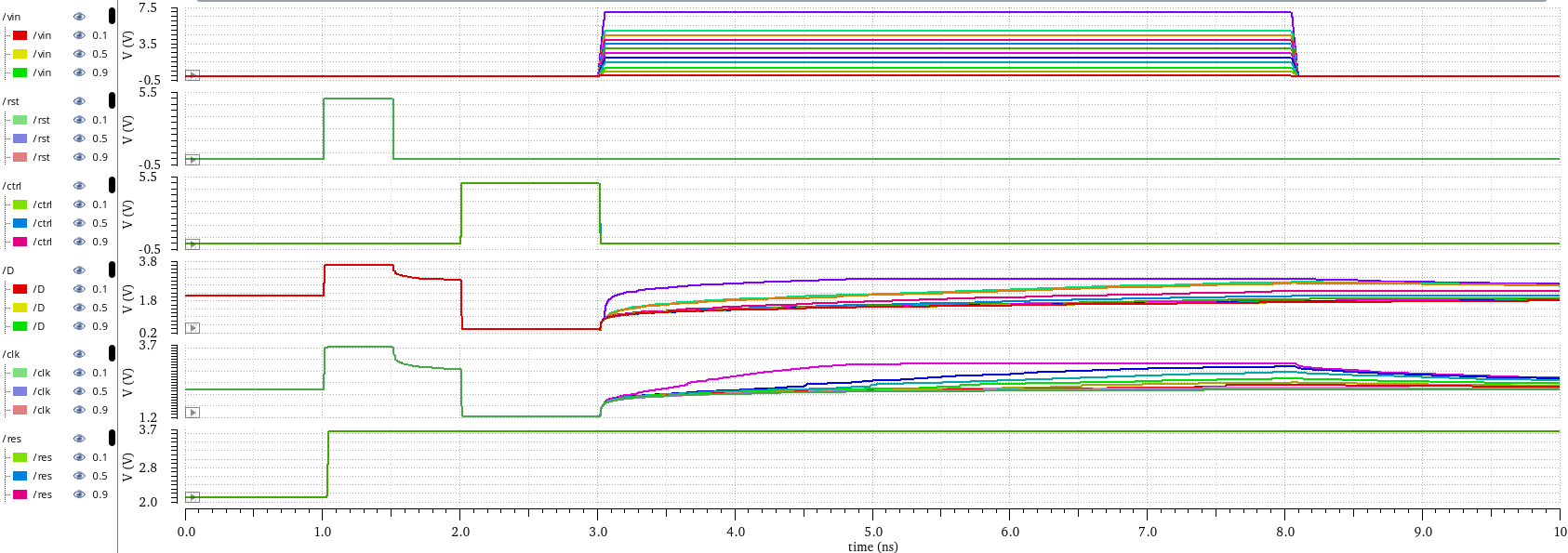}
\caption{Supply voltage effect on output signals}
\label{fig4}
\end{figure}
\begin{figure}[htbp]
\centering
\includegraphics[width=0.48\textwidth]{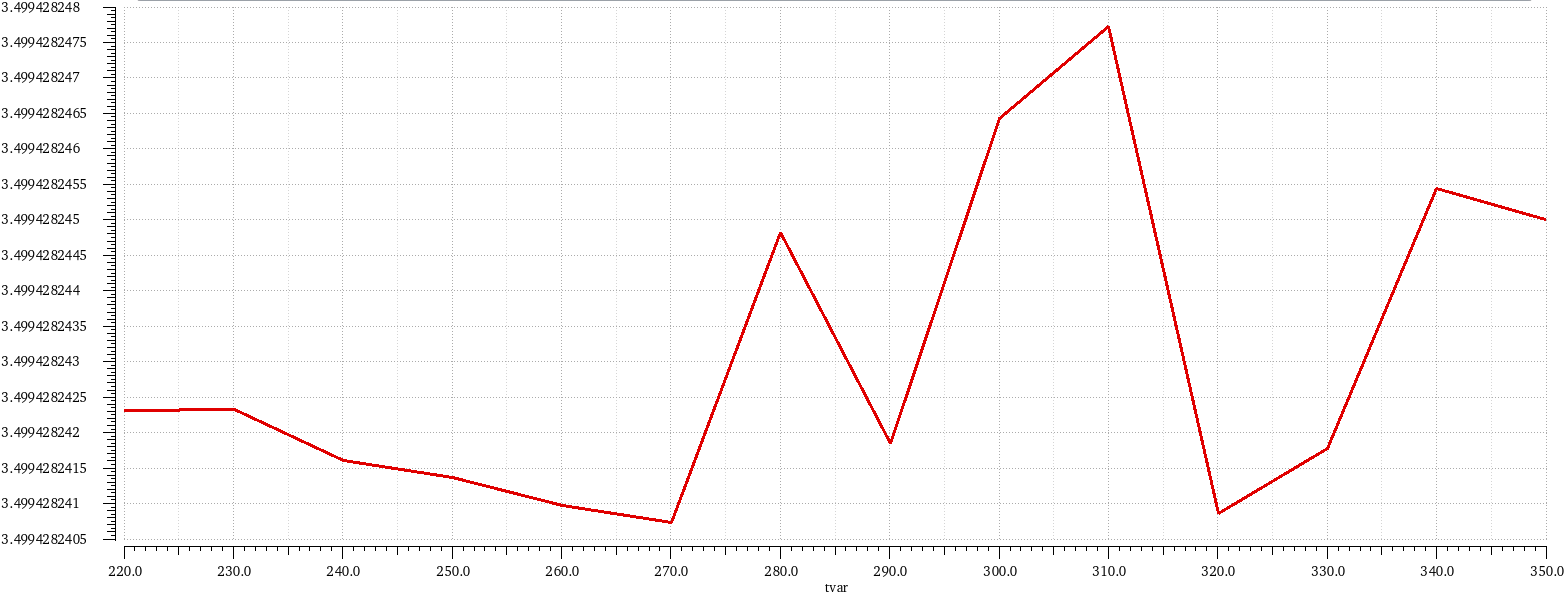}
\caption{Analog response value VS temperature (memristor)}
\label{fig5}
\end{figure}

Using the formula, we estimated the response value from plots. For the reliability test, the effects of temperature and supply voltage on the output response are shown. The variation of analog voltage values with these parameters is plotted. The effect of temperature has been considered for both CMOS- and memristor-based designs. Figures 4, 5, and 6 show the supply voltage effect on output signals, the analog response value vs. temperature (memristor), and the temperature effect on output signals (CMOS), respectively.
\begin{figure}[htbp]
\centering
\includegraphics[width=0.48\textwidth]{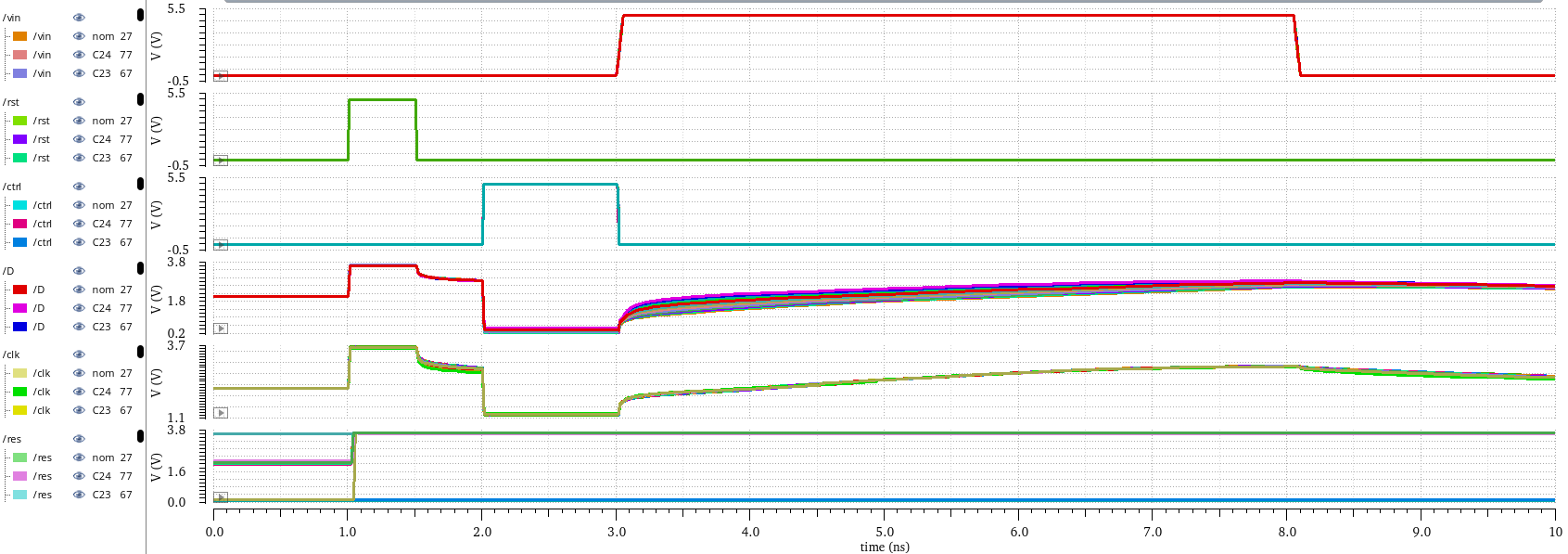}
\caption{Temperature (CMOS) effect on output signals}
\label{fig6}
\end{figure}
It is evident from Figures 4 and 6 that a delay between the D input and the clock occurs due to variations in supply voltage and CMOS temperature. Although several similar plots exhibit this behavior, one representative example is shown in Figure 5 to illustrate how the analog response value from the D flip-flop changes under these variations for direct value prediction.
\begin{figure}[htbp]
\centering
\includegraphics[width=0.48\textwidth]{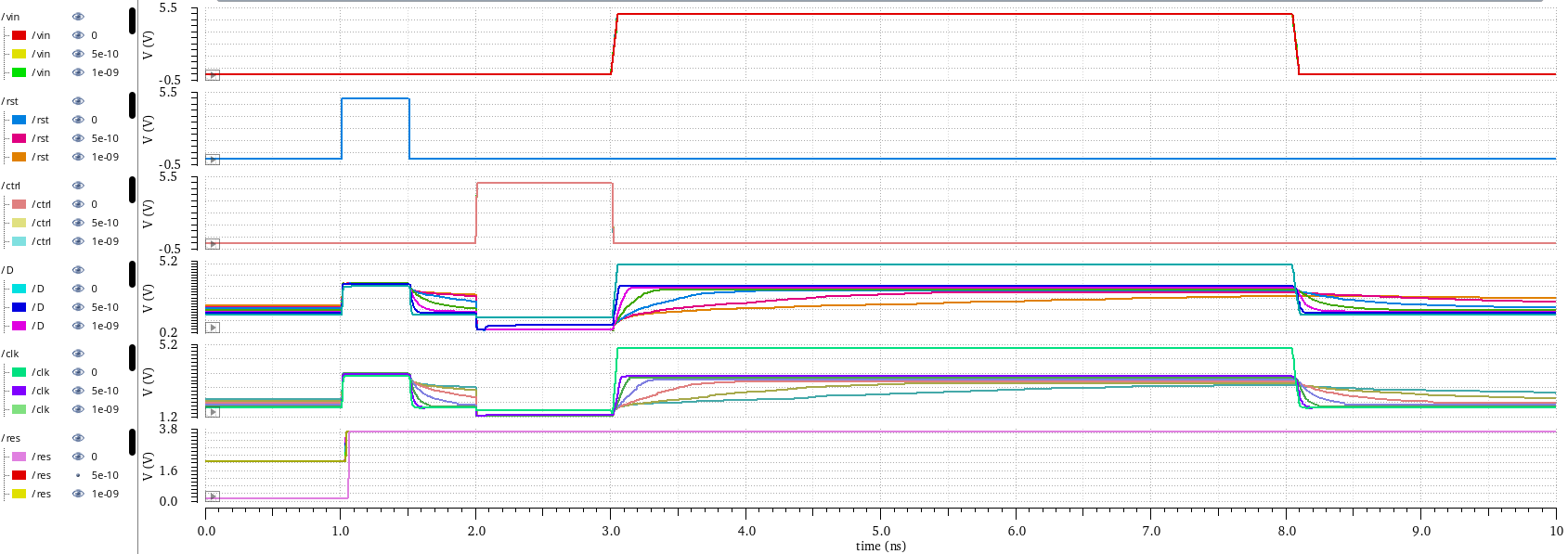}
\caption{Filament length effect on output signals}
\label{fig7}
\end{figure}
\begin{figure}[htbp]
\centering
\includegraphics[width=0.48\textwidth]{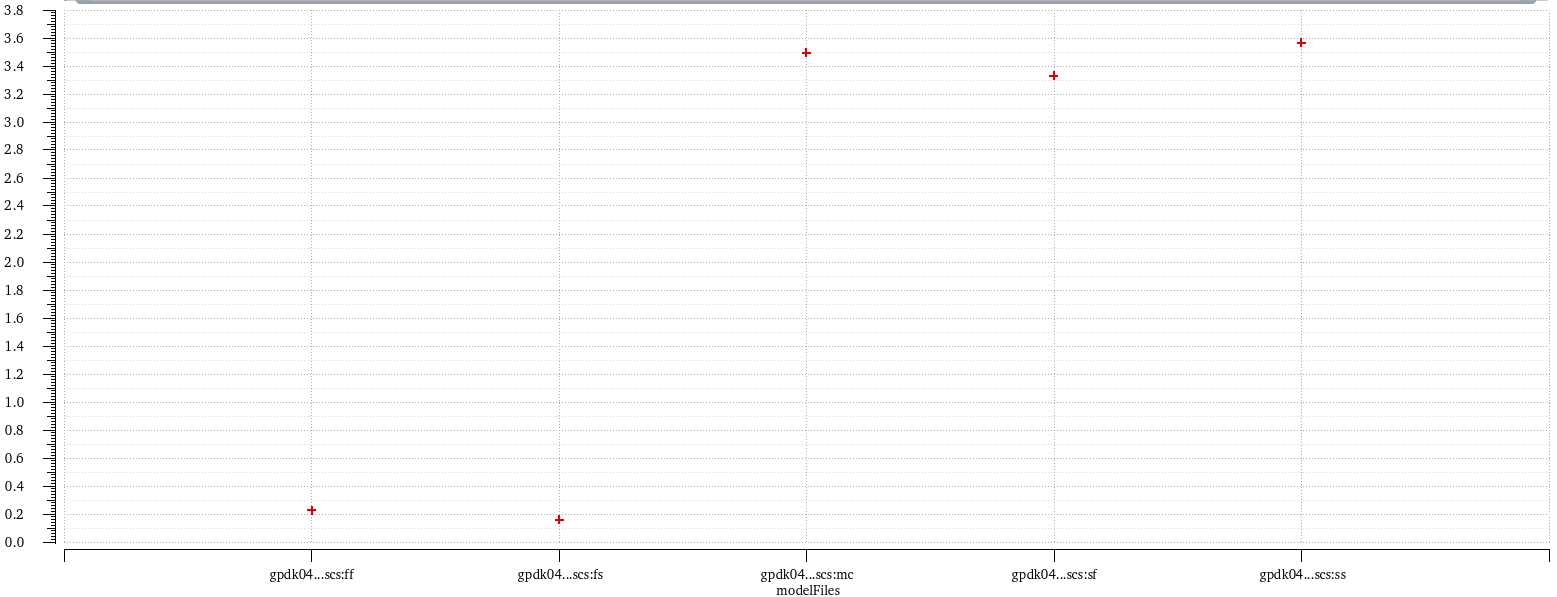}
\caption{Analog response value VS CMOS process corners}
\label{fig8}
\end{figure}

Response values for uniqueness were calculated using the same process. Figures 7 and 8 illustrate the effect of filament width on output signals and the variation of analog response values across CMOS process corners, respectively. Additional parameters such as hopping current density, $X_t$ effect, filament width, filament length, and doped region length were also considered. As shown in Figure 8, each CMOS process corner (e.g., ss, sf, fs, ff, etc.) produces a distinct analog response.
Similarly, for multiple-response analysis, responses were estimated by observing the output voltage while varying different parameters listed in Tables I and III.

To evaluate the statistical behavior of the proposed PUF configurations under process variations, Monte Carlo simulations were conducted using 350 samples. The analog response distributions were analyzed for both single-response and multiple-response PUF configurations, as illustrated in Figures 9 and 10, respectively. Histograms were plotted, where the x-axis denotes the response values and the y-axis indicates the corresponding frequency of occurrence.
\begin{figure}[htbp]
\centering
\includegraphics[width=0.48\textwidth]{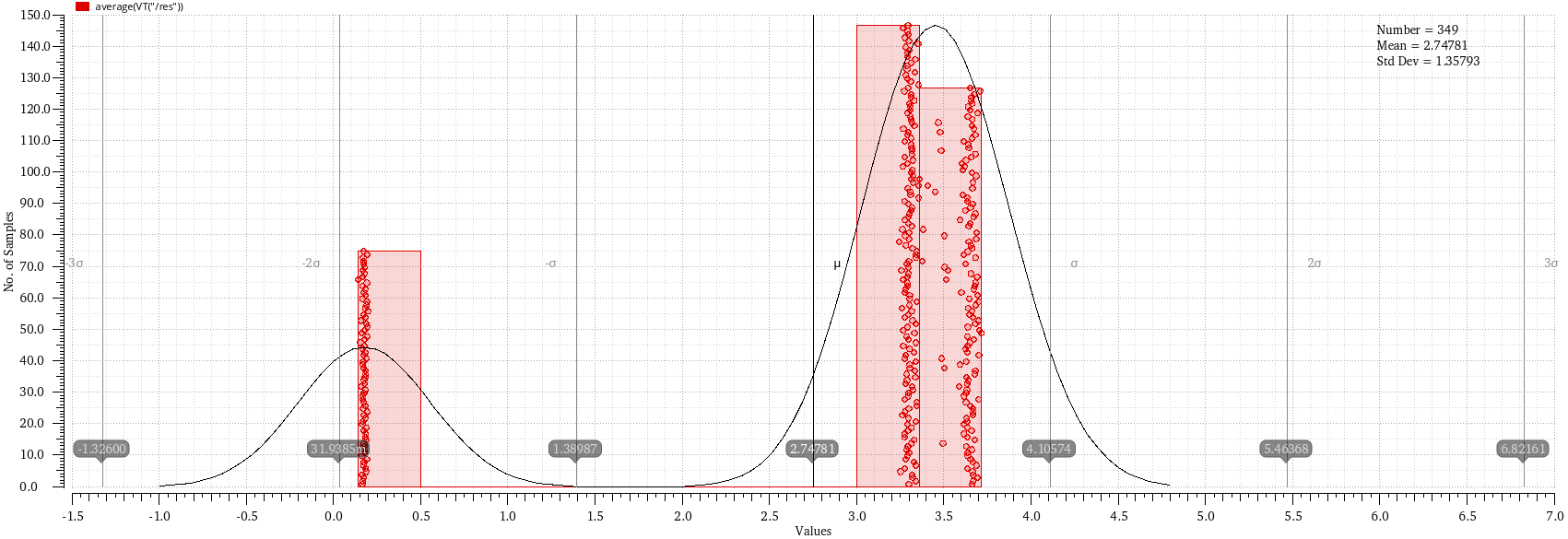}
\caption{Distribution of average analog response using Monte Carlo Simulation for 350 samples under CMOS process variation and mismatch}
\label{fig9}
\end{figure}
\begin{figure}[htbp]
\centering
\includegraphics[width=0.48\textwidth]{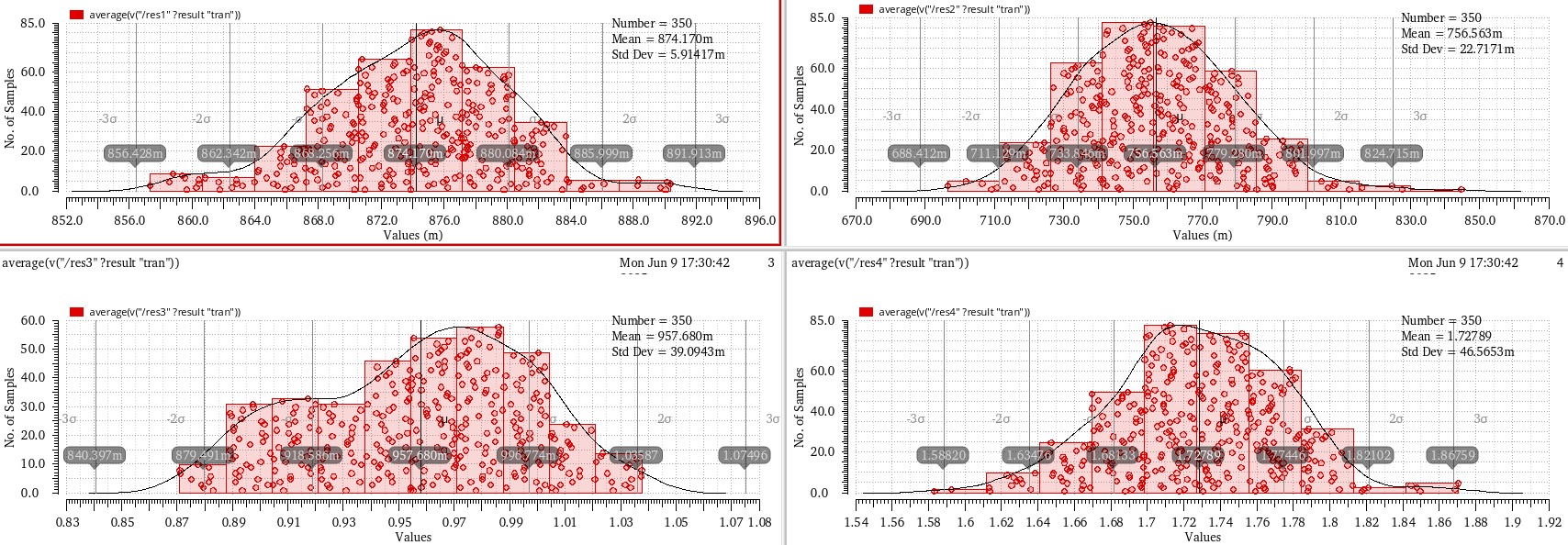}
\caption{Distribution of all average analog responses using Monte Carlo Simulation for 350 samples under CMOS process variation and mismatch}
\label{fig10}
\end{figure}

For the single-response PUF, the analog output exhibited a mean value of 2.75 with a standard deviation of 1.36, indicating a moderate spread in response values due to inherent circuit variations. In the case of the multiple-response PUF configuration, four distinct responses were analyzed. The mean values of these responses were 0.87, 0.76, 0.96, and 1.73, while the corresponding standard deviations were 0.006, 0.023, 0.039, and 0.047, respectively. These results suggest that the multiple-response configuration produces more tightly clustered outputs with lower variability compared to the single-response case, potentially contributing to improved response stability and distinguishability across different instances.

\section{Discussion}
The response values were estimated from the plots. Using Equation (1) for uniqueness and Equations (2) and (3) for reliability, the results are summarized in Table IV. Table V presents a comparison between this work and previously published studies.

\begin{table}[hbtp]
\renewcommand{\arraystretch}{1.6}
\setlength{\tabcolsep}{3pt}
\scriptsize
\centering
\caption{Reliability and Uniqueness of Proposed PUF}
\label{tab4:reliability_uniqueness}
\begin{tabular}{|c|c|c|c|c|}
\hline
\textbf{} & \textbf{CMOS} & \textbf{Memristor} & \textbf{Memristor} & \textbf{Memristor}\\ \textbf{} & \textbf{} & \textbf{(1res\_1T1M)} & \textbf{(4res\_1T1M\_2Stage)} & \textbf{(4res\_1T1M\_4Stage)} \\
\hline
\textbf{Size} & 16 bits & 8 bits & 8 bits & 16 bits \\
\hline
\textbf{Reliability (\%)} & 76.19 & 88.64 & 98.78 & 99.38 \\
\hline
\textbf{Uniqueness (\%)} & 45.98 & 50.13 & 12.61 & 12.47 \\
\hline
\end{tabular}
\end{table}

\begin{table*}[hbtp]
\renewcommand{\arraystretch}{1.6}
\setlength{\tabcolsep}{7pt}
\scriptsize
\centering
\caption{Comparison between this work and other papers}
\label{tab5:reliability_uniqueness}
\begin{tabular}{|c|c|c|c|c|c|c|c|c|}
\hline
\textbf{} & \textbf{} & \textbf{Technology} & \textbf{} & \textbf{Memristor} & \textbf{} & \textbf{No. of} & \textbf{Uniqueness} & \textbf{Reliability} \\ \textbf{} & \textbf{Year} & \textbf{Design} & \textbf{Topology} & \textbf{model} & \textbf{CRPs} & \textbf{response} & \textbf{(\%)} & \textbf{(\%)} \\ \textbf{} & \textbf{} & \textbf{Environment} & \textbf{} & \textbf{} & \textbf{} & \textbf{bits} & \textbf{} & \textbf{} \\
\hline
\textbf{This work} & \textbf{} & \textbf{Cadence} & \textbf{1res\_1T1M} & \textbf{} & \textbf{} & \textbf{} & \textbf{} & \textbf{} \\ \textbf{(Single} & \textbf{2025} & \textbf{gpdk45} & \textbf{APUF} & \textbf{Stanford} & $\mathbf{2^8}$ & \textbf{1} & \textbf{50.13} & \textbf{88.64} \\ \textbf{response)} & \textbf{} & \textbf{} & \textbf{} & \textbf{} & \textbf{} & \textbf{} & \textbf{} & \textbf{} \\
\hline
\textbf{This work} & \textbf{} & \textbf{Cadence} & \textbf{4res\_1T1M} & \textbf{} & \textbf{} & \textbf{} & \textbf{} & \textbf{} \\ \textbf{(Multiple} & \textbf{2025} & \textbf{gpdk45} & \textbf{4Stage} & \textbf{Stanford} & $\mathbf{2^{16}}$ & \textbf{4} & \textbf{12.47} & \textbf{99.38} \\ \textbf{responses)} & \textbf{} & \textbf{} & \textbf{APUF} & \textbf{} & \textbf{} & \textbf{} & \textbf{} & \textbf{} \\
\hline
\textbf{Chatterjee et} & 2016 & SPICE & 1res\_1T1M & Biolek & ${2^8}$ & 1 & 51.06 & 99.25 \\ \textbf{al. [9]} & & 45nm & APUF & & & & & \\
\hline
\textbf{Teo et al. [11]} & 2019 & SilTerra & 4res\_1T1M & Biolek & ${2^{16}}$ & 4 & 49.338 & - \\ \textbf{} & & 180nm & 4Stage APUF& & & & & \\
\hline
\textbf{Ge et al. [15]} & 2020 & Altera & CPP-APUF & - & 170K & 1 & 51.06 & 99.67 \\
\hline
\textbf{Mursi et al.} & 2021 & Artix-7 & CDC-XPUF & - & 300M & 1 & 17 & 97.5\\ \textbf{[16]} & & & & & & & & \\
\hline
 \textbf{Wisiol et al.} & & & & & & & 46.15 & \\ \textbf{[17]} & 2022 & - & LP-PUF & - & $2^{(N/M)}$ & 1 & \textbf{(Suh et al.} & 87.65 \\ \textbf{} & & & & & & & \textbf{[23])} & \\
\hline
 &  & - & SBC-PUF & Linear-ion & $2^N (N/2)^2$ & 1 & 50 & - \\
\cline{3-9}
\textbf{\centering Sun et al. [18]} & {2021}  & - & TBC-PUF & Linear-ion & $2^N$ & 1 & 50.1 & - \\
\cline{3-9}
 &  & - & MA-PUF & Linear-ion & $2^{(N+2)}$ & 1 & 50.6 & - \\
\hline
\textbf{Cui et al. [19]} & 2020 & Artix-7, & MMPUF & - & $2^{N}$ & 1 & 40.6 & 95 \\ \textbf{} & & Kintex-7 & & & & & & \\
\hline
\end{tabular}
\end{table*}

The results indicate that memristor-based PUFs exhibit superior reliability, achieving 98.78\% for the two-stage design and 99.38\% for the four-stage design, compared to 76.19\% for CMOS-based implementations, particularly in multi-response configurations. However, the uniqueness of memristor-based designs is slightly lower, with values of 12.61\% for the two-stage and 12.47\% for the four-stage configurations. Furthermore, single response PUF exhibits about 88.64\% of reliability and 50.13\% of uniqueness. These findings suggest that memristor-based architectures offer a promising alternative for secure authentication applications, although further optimization is required to achieve a balance between uniqueness and reliability.

This study has been compared with existing research based on several key factors, including the technological design environment, circuit topology, memristor model, challenge-response pairs (CRPs), number of response bits, and key PUF metrics such as uniqueness and reliability. These comparisons are summarized in Table V.
\section{Conclusion}
This paper emphasizes the potential of memristor-based Physically Unclonable Functions (PUFs) compared to conventional CMOS designs, particularly in providing greater reliability in multi-response setups. Although the uniqueness is somewhat lower, this can be enhanced through further refinements in design. The research investigates how circuit architecture, technology design, and memristor models influence crucial PUF metrics such as uniqueness and reliability. An innovative approach involves placing arbiters at various points along delay paths to increase resistance to attacks, and utilizes SR NAND latches instead of D flip-flops to minimize overhead. The Stanford memristor model, which features filament-based resistive switching and variability, is vital in enhancing randomness and unpredictability. In summary, the proposed design bolsters hardware-level security, presenting a scalable and efficient solution for secure hardware applications.

\vspace{12pt}

\end{document}